\documentclass[fleqn,usenatbib]{mnras}
\usepackage{mathptmx}
\usepackage[T1]{fontenc}
\usepackage{ae,aecompl}

\usepackage{graphicx}	
\usepackage{amsmath}	
\usepackage{amssymb}	
\usepackage{tabularx}
\usepackage{subfigure}
\usepackage[flushleft]{threeparttable}
\usepackage[a4paper]{geometry}%
\usepackage{hyperref}

\newcommand{\Msun}{~M_\odot}

\newcommand{\cmc}{\rm ~cm^{-3}}
\newcommand{\kms}{\rm ~km~s^{-1}}

\newcommand{\ml}{~\Msun ~\rm yr^{-1}}

\title[Shock evolution in non-radiative SNRs]{SHOCK EVOLUTION IN NON-RADIATIVE SUPERNOVA REMNANTS}

\author[Xiaping Tang and Roger A. Chevalier]{
Xiaping Tang,$^{1}$\thanks{E-mail: xt5uv@mpa-garching.mpg.de}
Roger A. Chevalier$^{2}$
\\
$^{1}$Max Planck Institute for Astrophysics,
Karl-Schwarzschild-Str. 1,
D-85741 Garching, Germany\\
$^{2}$Department of Astronomy, University of Virginia, P.O. Box 400325,
Charlottesville, VA 22904-4325, USA
}

\date{Accepted XXX. Received YYY; in original form ZZZ}

\pubyear{2016}

\begin{document}
\label{firstpage}
\pagerange{\pageref{firstpage}--\pageref{lastpage}}
\maketitle

\begin{abstract}
We present a new analytical approach to derive approximate solutions describing the shock evolution in non-radiative supernova remnants (SNRs). We focus on the study of the forward shock and contact discontinuity while application to the reverse shock is only discussed briefly. The spherical shock evolution of a SNR in both the interstellar medium with a constant density profile and a circumstellar medium with a wind density profile is investigated. We compared our new analytical solution with numerical simulations and found that a few percent accuracy is achieved.  For the evolution of the forward shock, we also compared our new solution to previous analytical models. In a uniform ambient medium, the accuracy of our analytical approximation is comparable to that in \cite{TM99}. In a wind density profile medium, our solution performs better than that in \cite{Micelotta16}, especially when the ejecta envelope has a steep density profile.  The new solution is significantly simplified compared to previous analytical models, as it only depends on the asymptotic behaviors of the remnant during its evolution.
\end{abstract}




\section{Introduction}
In this work, we derive simple analytical formulae characterizing the shock evolution in a non-radiative supernova remnant (SNR). In order to obtain an analytical solution, we constrain our discussion to a simple situation: spherical expansion in a smooth medium (no clouds) with negligible external thermal pressure. Thermal conduction, magnetic fields and acceleration of cosmic ray particles are also neglected for simplicity. The possible extension of the current model to more complicated situations will be studied in future work. 

The work presented here focuses on the evolution of a remnant in the post supernova phase. A point explosion with an ejecta mass of $M_{ej}$ and total energy of $E_{SN}$ is assumed as our initial condition. The energetic ejecta released in the supernova explosion  drive a blast wave into the surrounding ambient medium, which is assumed to have a power law density profile with index $s$, i.e. $\rho_a \propto R^{-s}$. During the interaction between ejecta and ambient medium, both a forward shock into the surrounding medium with radius $R_b$ and a reverse shock into the expanding ejecta with radius $R_r$ are generated. The interface between the ejecta and the ambient medium is the shock contact discontinuity (CD). Its radius is defined as $R_c$. 

At early times, when the ejecta mass $M_{ej}\gg M_{sw}$, the swept up  mass, the evolution of the remnant can follow two different evolutionary tracks depending on the spatial density distribution of the ejecta envelope. If the ejecta envelope has a shallow density profile $\rho \propto R^{-n}$ with power law index $n<5$, the early evolution of a SNR is characterized by the free expansion (FE) of the ejecta, with a narrow outer shocked region. In the FE solution, the CD  expands freely with a constant velocity while the forward shock follows a similarity solution  \citep{Parker63,HS84}. Due to the accumulation of shocked ambient medium ahead of the CD, it is found that $R_b=q_{b}R_c$ where $q_{b}>1$ is a dimensionless constant. 
If the ejecta envelope has a steep density profile $\rho \propto R^{-n}$ with power law index $n>5$, the early evolution of a SNR is instead described by the  self similar driven wave (SSDW) solution \citep{Chevalier82,nadezhin85}. In the SSDW solution, $R_c\propto t^{(n-3)/(n-s)}$ based on dimensional analysis while $R_b=q_{b}R_c$ and $R_r=q_{r}R_c$ where $q_{b}$ and $q_{r}$ are dimensionless constants. 

As the blast wave expands, $M_{sw}$ gradually increases with time and eventually becomes dynamically important. When $M_{sw} \gg M_{ej}$, the expanding SNR has already lost the memory of ejecta mass $M_{ej}$ and starts to follow the self similar Sedov-Taylor (ST) solution \citep{Taylor46,Sedov59} in which $R_b\propto t^{2/(5-s)}$. 


According to the asymptotic behaviors described above at $M_{ej}\gg M_{sw}$ and $M_{ej}\ll M_{sw}$, \citep[][hereafter TM99]{TM99}  derive analytic approximations for the evolution of the forward shock and reverse shock in a non-radiative SNR with further dynamical considerations. In TM99, the solution for the forward shock contains two parts: {\it a general ED solution} for the ED phase and {\it a general ST solution} for the ST phase. The transition time $t_{ST}$ defined in TM99, which separates the {\it general ED solution} from the {\it general ST solution}, is slightly different from the time when $M_{ej}=M_{sw}$ and in many cases is obtained through fitting  numerical simulations. The {\it general ED solution} asymptotically approaches the FE solution when $n<5$ and the SSDW solution when $n>5$ as $t\rightarrow 0$, and is extended to finite $t$ by assuming the pressure behind the blast wave is proportional to that behind the reverse shock. It has two different forms depending on whether the reverse shock is in the envelope or the core of the ejecta. The {\it general ST solution} approaches the ST solution as $t\rightarrow \infty$ and equals to the value of the {\it general ED solution} at $t_{ST}$. It has the form of an offset power law and is designed to smoothly connect the {\it general ED solution} and the ST solution.  
The solution for the reverse shock in TM99 also contains two parts. In the ED phase, the reverse shock position is derived by assuming the reverse shock radius is proportional to the forward shock radius. In the ST phase, the reverse shock is described by a solution with constant acceleration in the unshocked ejecta frame. 

TM99 applied their method to a power law density ambient medium with a focus on the uniform medium and a brief discussion about the wind density profile. \cite{L&H03}, \cite{HL12} and \cite{Micelotta16} then studied the wind density profile in more detail with the method described in TM99. However, their solutions are not compared to numerical simulations. \cite{L&H03} and \cite{HL12} also presented analytical approximate formulae for the fitting coefficients used in the TM99 solution. Since all these analytical solutions \citep{TM99,L&H03,HL12,Micelotta16} are based on the method of TM99, from here on we refer to the TM model as the combination of the above solutions.

Here we present a new analytical method to derive approximate solutions describing the shock evolution in a SNR from the ED phase to the ST phase. The method is based on dimensional analysis and depends on only the asymptotic behaviors of the remnant, i.e. the FE solution and the SSDW solution for $M_{sw}\ll M_{ej}$ and the ST solution for $M_{sw}\gg M_{ej}$.  Because no further assumptions about the dynamical structure of the remnant are required as in the TM model, the analytical approximations discussed here are much simpler than the TM model solutions. The method presented here could potentially be extended to other problems involving the transition between two adjacent asymptotic limits.

In Section \ref{sec:method}, we develop the analytical approach used to derive the approximate solutions. Then we use the new method to study the evolution of the forward shock and CD in a non-radiative SNR. Analytical approximations for both ejecta envelope with a shallow density profile $n<5$ and a steep density profile $n>5$ are investigated in detail. In Section \ref{sec:comparison}, we summarize the analytical approximations for both forward shock and CD, and then compare them with numerical simulations. For the forward shock, we also compare our new solutions with those from the TM model. We focus on two particularly interesting cases: SNR evolution in the interstellar medium with a constant density profile and SNR evolution in circumstellar material with a wind density profile. A reader who is only interested in the final expressions of the analytical approximations can go directly to this section. In Section \ref{sec:RS},  application of our new method to the reverse shock is discussed briefly. A final discussion and summary are in Section \ref{sec:DS}.
\section{BASIC METHOD}{\label{sec:method}}
\subsection{Dimensional analysis}
Based on the $\Pi$ theorem \cite[see, e.g., Chapter 1 of][]{Barenblatt96}, a physical relation involving $k+m$ physical variables with $k$ independent physical dimensions can be simplified into a physical relation with only $m$ independent dimensionless quantities. In other words, an equation
\begin{equation}
f(a_1,...,a_k,...,a_{k+m})=0
\end{equation} 
involving $k$ independent physical dimensions is equivalent to the following simplified equation
\begin{equation}
F(\Pi_1,...,\Pi_m)=0,
\label{dim_general}
\end{equation}
where $\Pi_1, ...,\Pi_m$ are independent dimensionless quantities built by a combination of $a_1,..., a_{k+m}$.
If $m$ happens to be 1, eq. (\ref{dim_general}) then becomes $F(\Pi)=0$ and has a trivial solution $\Pi =C$, where $C$ is a constant. When $C \neq 0$, according to dimensional analysis, a self similar solution of the first kind exists for the problem. The evolution of such a system is characterized by the invariant dimensionless quantity $\Pi = C$. 

The shock evolution in a non-radiative SNR under our simplified assumptions involves 5 different dimensional physical variables: explosion energy $E_{SN}$, ejecta mass $M_{ej}$, ambient medium density $\rho_a$, remnant age $t$, and blast wave radius $R_b$ (or CD radius $R_c$ and reverse shock radius $R_r$ depending on your interest). In this Section, we focus on the study of the forward shock while an approximate solution for the CD is presented at the end of this section. The reverse shock is discussed briefly in Section \ref{sec:RS}. 

Our primary initial goal is to derive an analytical approximation for the physical relation
\begin{equation}
f_b(E_{SN},M_{ej},\rho_a,t,R_b)=0.
\label{eq_general}
\end{equation} 
The problem has 3 independent physical dimensions: length, time and mass. According to the $\Pi$ theorem, eq. (\ref{eq_general}) is equivalent to the following relation 
\begin{equation}
F_b(\Pi_1,\Pi_2)=0,
\label{dim}
\end{equation}
where 
\begin{equation}
\Pi_1=\left(\frac{R_b}{t}\right)^2\left(\frac{M_{ej}}{E_{SN}}\right)
\end{equation}
and
\begin{equation}
\Pi_2=\frac{R_b^5\rho_a}{ E_{SN}t^2}
\end{equation} 
are the 2 independent dimensionless quantities available for our problem, 

Eq. (\ref{dim}) offers a complete description of the shock evolution in a non-radiative SNR. The exact form of eq. (\ref{dim})  must depend on physical considerations during the transition time and may not have a simple solution in the form of $R(t)$ or $t(R)$.  We instead seek analytical approximations for $F_b(\Pi_1,\Pi_2)$ which have a simple functional form and are consistent with numerical simulations within a few percent. The solution discussed below could be easily applied as a tool for more complicated problems involving the shock evolution in non-radiative SNRs.

\subsection{Characteristic scales}{\label{scale}}
Before we present our analytical solutions and compare them with numerical simulations, we first define the characteristic scales of the system to further simplify the expression. If we assume the  ambient medium has a power law density profile, i.e. $\rho_a(r) = \eta_s r^{-s}$, where $\eta_s$ is a constant. Then the characteristic length, time and mass of the system are as follows 
\begin{eqnarray}
M_{ch}&=&M_{ej},\\
R_{ch}&=&M_{ej}^{1/(3-s)}\eta_s^{-1/(3-s)},\\
t_{ch}&=&E_{SN}^{-1/2}M_{ej}^{(5-s)/2(3-s)}\eta_s^{-1/(3-s)}.
\end{eqnarray}
We denote the physical quantity $X$ in units of the corresponding characteristic scale as $X^*$, i.e. $X^*=X/M_{ch}^{x_1}R_{ch}^{x_2}t_{ch}^{x_3}$, where $x_1,x_2$ and $x_3$ are constants depending on the dimension of the quantity. In the rest of the paper, unless specifically noted, we  use the dimensionless quantity $X^*$
instead of $X$ throughout our discussion.

We are particularly interested in two situations: SNR evolution in the interstellar medium with a constant density profile ($s=0$) and SNR evolution in circumstellar matter with a wind density profile ($s=2$). For a uniform ambient medium, the characteristic radius and time are
\begin{equation}
R_{ch}=3.4~ {\rm pc}\left(\frac{M_{ej}}{\Msun}\right)^{1/3}\left( \frac{m_p\rm \cmc}{\eta_s}\right)^{1/3},
\end{equation}
and 
\begin{equation}
t_{ch}=473~ {\rm yr}\left(\frac{10^{51} {\rm erg}}{E_{SN}}\right)^{1/2}\left(\frac{M_{ej}}{\Msun}\right)^{5/6}\left(\frac{m_p {\rm \cmc}}{\eta_s}\right)^{1/3},
\end{equation}
where $m_p$ is the proton mass.
For a wind density profile, $\eta_s=\dot{M}_w/4\pi v_w$ where $\dot{M}_w$ is the mass loss rate and $v_w$ is the wind velocity. The characteristic radius and time now are
\begin{equation}
R_{ch}=12.9~ {\rm pc}\left(\frac{M_{ej}}{\Msun}\right) \left(\frac{10^{-5}\ml}{\dot{M}_w}\right) \left(\frac{v_w}{10 {\rm \kms}}\right)
\label{charact_R}
\end{equation}
and
\begin{equation*}
t_{ch}=1772~ {\rm yr}\left(\frac{10^{51} {\rm erg}}{E_{SN}}\right)^{1/2}\left(\frac{M_{ej}}{\Msun} \right)^{3/2}
\end{equation*}
\begin{equation}
\quad \,\times\left(\frac{10^{-5}\ml}{\dot{M}_w}\right) \left(\frac{v_w}{10 {\rm \kms}}\right).
\end{equation}

The dimensionless quantities $\Pi_1$ and $\Pi_2$ now simply become
\begin{equation}
\Pi_1=\left(\frac{R_b^*}{t^*}\right)^2 \quad \mbox{and} \quad \Pi_2=\frac{R_b^{*5-s}}{ t^{*2}}
\end{equation}

\subsection{Asymptotic behavior of the forward shock}\label{AB}
In this subsection, we examine the asymptotic behaviors of the forward shock during the non-radiative evolution, which is essential to derive the analytical approximation. The discussion here about the asymptotic behavior of a SNR is general and in principle could be extended to more complicated situations. But to obtain an explicit expression of the asymptotic solution,  we have to make some assumptions about the density distribution in the ejecta and ambient medium. In this paper, we apply the same density distribution as that in TM99, which is presented in detail in Appendix \ref{App:basic_params}. Basically, the ejecta have a flat core, with core radius to ejecta radius ratio $w_{core}$, and a power law envelope with index $n$, while the ambient medium is assumed to have a power law profile with index $s$.  

When $t\rightarrow \infty$, the remnant approaches the ST solution and the blast wave radius 
$R_b^*=(\xi t^{*2})^{1/(5-s)}$, i.e.
\begin{equation}
F_b(\Pi_1,\Pi_2)(t\rightarrow \infty)=F_b(\Pi_2)=\Pi_2-\xi=0,
\label{ST}
\end{equation}
where $\xi$ is a dimensionless constant depending on the density structure of the ambient medium $\rho_a$.

When $t\rightarrow 0$, the asymptotic behavior of SNRs becomes slightly complicated as we now have two different situations. If the ejecta have a shallow envelope in density with power law index $n<5$, the remnant simply follows the FE solution and the forward shock radius $R^*_b=q_bR^*_c=q_b\lambda_c t^*$ where $\lambda_{c}$ is a dimensionless constant depending on the density structure of the ejecta $\rho_{ej}$ \citep{Parker63,HS84}. We define $\lambda_b=q_b\lambda_c$ and then the blast wave radius $R^*_{b}=\lambda_{b} t^*$ as $t\rightarrow 0$, i.e.
\begin{equation}
F_b(\Pi_1,\Pi_2)(t\rightarrow 0)=F_b(\Pi_1)=\Pi_1-\lambda_{b}^2=0.
\label{FE}
\end{equation} 

If the ejecta have a steep envelope with power law index $n>5$, the remnant instead asymptotically approaches the SSDW solution and the forward shock radius $R_b^*=\zeta_bt^{*(n-3)/(n-s)}$  \citep{Chevalier82} as $t\rightarrow 0$, i.e.
 \begin{equation}
F_b(\Pi_1,\Pi_2)(t\rightarrow 0)=\Pi_1^{(n-5)/2(n-s)}\Pi_2^{1/(n-s)}-\zeta_b=0,
 \label{SSDW}
 \end{equation}
where $\zeta_b$ is a dimensionless constant depending on the density structure of both the ejecta $\rho_{ej}$ and the ambient medium $\rho_a$.

In Appendix \ref{App:basic_params}, we derive the dimensionless constants $\lambda_b(n)$, $\xi(s)$ and $\zeta_b(n,s)$ based on the density profile assumed in TM99. The resulting expressions are summarized in Table \ref{forwardshock}. According to the above discussion, the ST solution can be considered as the asymptotic solution of the general equation $F_b(\Pi_1,\Pi_2)=0$ in the limit $t\rightarrow \infty$ while the FE solution and SSDW solution behave like the asymptotic solution of equation $F_b(\Pi_1,\Pi_2)=0$ in the limit $t\rightarrow 0$. No matter what functional form of approximation we choose for eq. (\ref{dim}), it must satisfy the asymptotic limits described in eqs. (\ref{ST}), (\ref{FE}) and (\ref{SSDW}).

\subsection{Analytical approximation for the forward shock }
The problem of deriving analytical solutions to the physical relation in eq. (\ref{eq_general})  is now simplified to the problem of finding approximate solutions for eq. (\ref{dim}) under the boundary conditions eqs. (\ref{ST}), (\ref{FE}) and (\ref{SSDW}). The primary goal of this subsection is to derive  analytical approximations for eq. (\ref{dim}) which have simple functional forms and satisfy the boundary conditions discussed before. More importantly, as we will show in the following section, the analytical approximations discussed here are consistent with numerical simulations within a few percent accuracy. 

In this work, we focus on an analytical approximation with the following polynomial form
\begin{equation}
F_b(\Pi_1,\Pi_2)=\left(\frac{\Pi_1}{\lambda_b^2}\right)^\alpha+\left(\frac{\Pi_2}{\xi}\right)^\beta-1=0\quad \mbox{for}\,\, n<5
\label{dim_special}
\end{equation}
and
 \begin{eqnarray}
F_b(\Pi_1,\Pi_2)&=&\left(\frac{\Pi_1^{(n-5)/2(n-s)}\Pi_2^{1/(n-s)}}{\zeta_b}\right)^{\alpha}\nonumber \\
&+&\left(\frac{\Pi_2}{\xi}\right)^{\beta}
-1=0 \,\, \mbox{for}\,\, n>5.
\label{dim_n>5}
\end{eqnarray}
When $\alpha$ and $\beta$ satisfy certain conditions, it can be shown that the above solutions naturally satisfy the boundary conditions eqs. (\ref{ST}), (\ref{FE}) and  (\ref{SSDW}). However such approximations are still complicated and do not always provide explicit expressions in the form of $R(t)$ or $t(R)$. If we further assume $2\alpha=(5-s)\beta>0$ for eq. (\ref{dim_special}) and $\alpha=(5-s)\beta>0$ for eq. (\ref{dim_n>5}), a simple analytical solution in the form of $R(t)$ can be derived easily. 
A simple analytical approximation in the form of $t(R)$ can also be obtained if we instead assume $\alpha=\beta>0$ for eq. (\ref{dim_special}) and $\alpha=2\beta (n-s)/(n-3)>0$ for eq. (\ref{dim_n>5}). $R(t)$ and $t(R)$ type solutions are two ways to approach the exact solution and approximate the evolution of the forward shock in a non-radiative SNR. Both of them are able to provide good fits to the simulations within a few percent accuracy, despite the fact that each of them may have its own advantage over certain parameter ranges. For simplicity, we will stick with one type of solution during our discussion. It is found that overall the $R(t)$ type solutions show slightly better performance than the $t(R)$  solutions. In the rest of this section, we will focus on the $R(t)$ type solution.

\begin{table}
\centering
\caption{Analytical approximation for the forward shock radius $R^*_{b}$}
\begin{threeparttable}
\begin{tabular}{c}
\hline\hline
$n<5$\\
\hline
\parbox{3cm}{
\begin{equation*}
R_b^*(t^*)=\left[\left(\lambda_b t^*\right)^{-2\alpha}+\left({\xi t^{*2}}\right)^{-2\alpha/(5-s)}\right]^{-1/2\alpha},
\end{equation*}}\\
\hline
$n>5$\\
\hline
\parbox{3cm}{
\begin{equation*}
R^*_b(t^*)=\left[\left(\zeta_b t^{*(n-3)/(n-s)}\right)^{-\alpha}+\left(\xi t^{*2}\right)^{-\alpha/(5-s)}\right]^{-1/\alpha}
\end{equation*}}\\
\hline
\parbox{3cm}{
\begin{equation*}
\lambda_c^2(n>3)=2w_{core}^{-2}\left(\frac{5-n}{3-n}\right)\left( \frac{w_{core}^{n-3}-n/3}{w_{core}^{n-5}-n/5} \right),
\end{equation*}
\begin{equation*}
\lambda_c^2(n<3)=2\left(\frac{5-n}{3-n}\right),
\end{equation*}
\begin{equation*}
^a\lambda_b=q_b\lambda_c \mbox{, where } q_b(s=0)=1.1 \mbox{ and } q_b(s=2)=1.19,
\end{equation*}
\begin{equation*}
^b\zeta_b =\left(\frac{R_1}{R_c}\right)\left(Af_0w_{core}^n\lambda_c^{n-3}\right)^{1/(n-s)},
\end{equation*}
\begin{equation*}
^c\xi(s=0)=2.026 \quad\mbox{and}\quad \xi(s=2)=3/2\pi,
\end{equation*}
}\\
\hline\hline
\end{tabular} 
\label{forwardshock}
\begin{tablenotes}
\small
\item[a] Reference \cite{Parker63} and \cite{HS84}.
\item[b] Exact values of $\zeta_b$ for $s=0$ and $s=2$ cases are presented in Table \ref{FSs0} and \ref{FSs2} respectively. See Appendix \ref{App:basic_params} for a detailed derivation.
\item[c] See eq. (\ref{xi}) in Appendix \ref{App:basic_params} for $\xi(s)$ with arbitrary $s$.
\item $\alpha$ is the free parameter in the model. The best fits $\alpha$ for $s=0$ and $s=2$  are recorded in Tables \ref{FSs0} and \ref{FSs2}, respectively.
\end{tablenotes}
\end{threeparttable}
\end{table}

\subsection{$n<5$ solution for the forward shock}
Assuming $2\alpha=(5-s)\beta>0$, eq. (\ref{dim_special}) now becomes 
\begin{equation}
\left(\frac{R_b^*}{\lambda_b t^*}\right)^{2\alpha}+\frac{R_b^{2\alpha}}{(\xi t^{*2})^{2\alpha/(5-s)}}=1.
\end{equation}
The solution of the equation can be expressed explicitly in the form   
\begin{equation}
R_b^*(t^*)=\left[\left(\lambda_b t^*\right)^{-2\alpha}+\left({\xi t^{*2}}\right)^{-2\alpha/(5-s)}\right]^{-1/2\alpha}.
\label{n<5}
\end{equation}

\begin{figure}
 \includegraphics[width=\columnwidth]{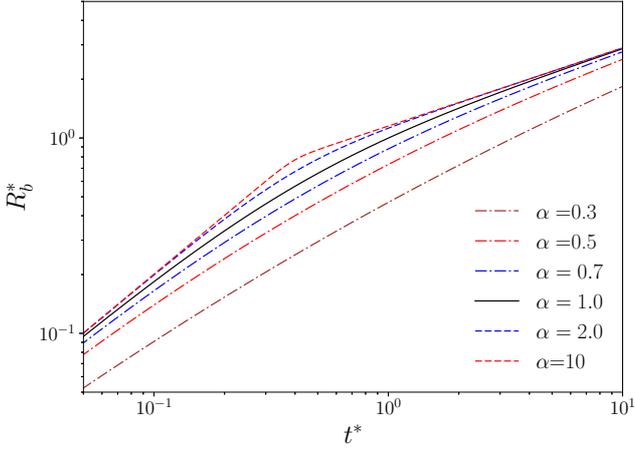} 
  \caption{Dimensionless forward shock radius $R_b^*$  as a function of the dimensionless time $t^*$ for various $\alpha$. The calculation is based on eq. (\ref{n<5}) with $n=0$ and $s=0$ .} 
    \label{various_alpha}
 \end{figure}

Equation (\ref{n<5}) with various $\alpha$ forms a group of curves representing different shapes of the transition from the FE solution to the ST solution. If we define a transition time $t^*_{tran}$ and radius $R^*_{tran}$ at which the two terms on the RHS of eq. (\ref{n<5}) are equal to each other, then we have
\begin{equation}
t^*_{tran}=\left(\frac{\xi}{\lambda_b^{5-s}}\right)^{1/(3-s)} \quad \mbox{and  }\quad
R^*_{tran}=2^{-1/2\alpha}\left(\frac{\xi}{\lambda_b^2}\right)^{1/(3-s)}.
\label{tran_n<5}
\end{equation}
$t^*_{tran}$ does not depend on $\alpha$ and can be considered as a critical time when the swept up mass becomes significant and dynamically important. The value of $R^*_{tran}$ does change with $\alpha$. When $\alpha\rightarrow\infty$, $R^*_{tran}\rightarrow (\xi/\lambda_b^2)^{1/(3-s)}$ and eq. (\ref{n<5}) represents an instantaneous transition from the FE solution to the ST solution. When $\alpha\rightarrow 0$, $R^*_{tran}\rightarrow 0$ and the equation instead characterizes a situation in which the system infinitely slowly and smoothly transits from one solution to another. In summary, when $\alpha$ varies from $0$ to $\infty$, the curve described by eq. (\ref{n<5}) changes from a slow and smooth transition to a break power law, as shown in Fig \ref{various_alpha}. It is expected that one of the curves in the group can approximate the evolution of a non-radiative SNR. Values of $t^*_{tran}$ and $R^*_{tran}$ calculated for various density structures are shown in Table \ref{FSs0} and \ref{FSs2} for $s=0$ and $s=2$, respectively. 

\subsection{$n>5$ solution for the forward shock}
Assuming $\alpha=(5-s)\beta>0$, eq. (\ref{dim_n>5}) then becomes
\begin{equation}
\left(\frac{R_b}{\zeta_bt^{(n-3)/(n-s)}}\right)^{\alpha}+\left(\frac{R_b^{*5-s}}{\xi t^{*2}}\right)^{\beta}=1.
\end{equation}
The solution to the equation can be expressed explicitly in the form 
\begin{equation}
R^*_b(t^*)=\left[\left(\zeta_b t^{*(n-3)/(n-s)}\right)^{-\alpha}+\left(\xi t^{*2}\right)^{-\alpha/(5-s)}\right]^{-1/\alpha}.
\label{n>5}
\end{equation}
Again we can define the transition time $t^*_{tran}$ and radius $R_{tran}^*$ between the SSDW solution and the ST solution:
\begin{equation}
t^*_{tran}=\left(\frac{\xi}{\zeta_b^{5-s} }\right)^{(n-s)/(n-5)(3-s)}
\end{equation}
and
\begin{equation}
R^*_{tran}=\frac{\zeta_b}{2^{1/\alpha}} t^{*(n-3)/(n-s)}_{tran}=\frac{\zeta_b}{2^{1/\alpha}} \left(\frac{\xi}{\zeta_b^{5-s} }\right)^{(n-3)/(n-5)(3-s)}.
\end{equation}
As in the $n<5$ case, $t^*_{tran}$ does not depend on $\alpha$ and characterizes the time when the swept up mass becomes significant and dynamically important. When we vary $\alpha$ we only manipulate the transition radius $R_{tran}^*$, which determines the smoothness of the transition.

\subsection{Contact Discontinuity}\label{sec:CD}
The time evolution of the CD radius in non-radiative SNRs has not been discussed before in the literature. Since the method developed here depends on only the asymptotic behavior of the remnant, in principle it can also be applied to the evolution of the CD radius $R_c^*$. The asymptotic behavior of $R_c^*$ at early times, when $t^*\rightarrow 0$, is simply the FE solution 
\begin{equation}
R_c^*=\lambda_c t^*  \quad \mbox{for}\quad n<5
\end{equation}
and the SSDW solution 
\begin{equation}
R_c^*=\zeta_c t^{*(n-3)/(n-s)}\quad \mbox{for}\quad n>5.
\end{equation} 
$\lambda_c$ and $\zeta_c$ are constants that depend on the density profile of ejecta and ambient medium. $\lambda_c$ is related to $\lambda_b$ \citep{HS84}  while $\zeta_c$ is proportional to $\zeta_b$ \citep{Chevalier82}. The detailed derivation of $\lambda_c$ and $\zeta_c$ is presented in Appendix \ref{App:basic_params}. The asymptotic behavior of the CD as $t^*\rightarrow \infty$, however, is not very clear at this point. If we assume the asymptotic behavior of the CD at $t^*\rightarrow \infty$ can be described by a simple power law relation $ct^{*b}$, where $c$ and $b$ are constants, then following the same spirit as for the forward shock we  obtain the following approximation for $R^*_c$:
\begin{equation}
\left(\frac{R^*_c}{\lambda_c t^*}\right)^\alpha+\left(\frac{R^*_c}{c t^{*b}}\right)^\alpha=1 \quad \mbox{for}\quad n>5
\end{equation}
and
\begin{equation}
\left(\frac{R^*_c}{\zeta_ct^{*(n-3)/(n-s)}}\right)^\alpha+\left(\frac{R^*_c}{c t^{*b}}\right)^\alpha=1 \quad \mbox{for}\quad n>5.
\end{equation}
Now $R^*_c$ has a simple analytical solution
\begin{equation}
R^*_c=\left[(\lambda_c t^*)^{-\alpha}+(c t^{*b})^{-\alpha}\right]^{-1/\alpha}\quad \mbox{for}\quad n<5,
\label{CDn<5}
\end{equation}
and
\begin{equation}
R^*_c=\left[(\zeta_ct^{*(n-3)/(n-s)})^{-\alpha}+(c t^{*b})^{-\alpha}\right]^{-1/\alpha}\quad \mbox{for}\quad n>5.
\label{CDn>5}
\end{equation}
In section \ref{sec:comparison}, we will show that the above solutions are able to provide good fits to numerical simulations within a few percent accuracy.

\begin{table}
\centering
\caption{Analytical approximation for the CD radius $R^*_{c}$}
\begin{threeparttable}
\begin{tabular}{c}
\hline\hline
$n<5$\\
\hline
\parbox{3cm}{
\begin{equation*}
R^*_c=\left[\left(\lambda_c t^*\right)^{-\alpha}+\left(c t^{*b}\right)^{-\alpha}\right]^{-1/\alpha},\quad \quad
\end{equation*}}\\
\hline
$n>5$\\
\hline
\parbox{3cm}{
\begin{equation*}
R^*_c=\left[\left(\zeta_ct^{*(n-3)/(n-s)}\right)^{-\alpha}+\left(c t^{*b}\right)^{-\alpha}\right]^{-1/\alpha},
\end{equation*}}\\
\hline
\parbox{3cm}{
\begin{equation*}
\lambda_c^2(n>3)=2w_{core}^{-2}\left(\frac{5-n}{3-n}\right)\left( \frac{w_{core}^{n-3}-n/3}{w_{core}^{n-5}-n/5} \right),
\end{equation*}
\begin{equation*}
\lambda_c^2(n<3)=2\left(\frac{5-n}{3-n}\right),
\end{equation*}
\begin{equation*}
^a\zeta_c =\left(Af_0w_{core}^n\lambda_c^{n-3}\right)^{1/(n-s)}
\end{equation*}}\\
\hline\hline
\end{tabular} 
\label{CD}
\begin{tablenotes}
\small
\item[a] Exact values of $\zeta_c$ for $s=0$ and $s=2$ cases are presented in Tables \ref{CDs0} and \ref{CDs2}, respectively. See Appendix \ref{App:basic_params} for a detailed derivation.
\item $\alpha$, $b$ and $c$ are free parameters of the model. Their best fit values for  the $s=0$ and $s=2$ cases are recorded in Table \ref{CDs0} and \ref{CDs2}, respectively.
\end{tablenotes}
\end{threeparttable}
\end{table}

\section{COMPARISON TO NUMERICAL SIMULATIONS AND THE TM MODEL}{\label{sec:comparison}}
In this section, we compare our analytical approximations for the forward shock radius $R_b^*$ and the CD radius $R_c^*$ to numerical simulations. For the forward shock, we also compare our new solution to the TM model results if available. Physical variables with the symbol $^*$ are dimensionless quantities in units of the characteristic scales defined in section \ref{scale}.

\subsection{Summary of our model}
Our analytical approximations for the forward shock and CD are summarized in Tables \ref{forwardshock} and \ref{CD}, respectively. $w_{core},n$ and $s$ are constants characterizing the initial density distribution in the ejecta and ambient medium. Basically, the ejecta have a flat core, with core radius to ejecta radius ratio $w_{core}$, and a power law envelope with index $n$, while the ambient medium is assumed to have a power law profile with index $s$. $\lambda_b$, $\lambda_c$, $\zeta_b$, $\zeta_c$ and $\xi$ are dimensionless constants describing the asymptotic behavior of the remnant that can be derived analytically. 
$\lambda_b$ and $\lambda_c$ correspond to the free expansion velocity of the forward shock and CD in the FE solution, respectively. $\zeta_b$ and $\zeta_c$ are the dimensionless constants for the forward shock and CD in the SSDW solution, respectively. $\xi$ is the dimensionless constant in the ST solution. For detailed definitions and derivations of the above parameters, see the discussion in Appendix \ref{App:basic_params}. 

For the forward shock, $\alpha$ is the only free parameter in our new analytical approximations, while for the CD we have three free parameters $\alpha$, $b$ and $c$ in our new solution. In this paper, we focus on two particularly interesting situations: SNR evolution in the interstellar medium with a constant density profile ($s=0$) and SNR evolution in circumstellar matter with a wind density profile ($s=2$). In each situation, we run numerical simulations with selected $n$ from 0 to 14 and then compare the simulation results with the analytical model to obtain the best fit free parameters. For the forward shock, the best fit $\alpha$ for the $s=0$ and $s=2$ cases with different $n$ are presented in Table \ref{FSs0} and \ref{FSs2} respectively. For the CD, the best fit $\alpha$, $b$ and $c$ for $s=0$ and $s=2$ cases with different $n$ are recorded in Table \ref{CDs0} and \ref{CDs2} respectively. 
Considering the uncertainty introduced by the code and the uncertainty in the input parameters like $\zeta_b$, $\zeta_c$, $\lambda_b$ and $\lambda_c$ which have only 2 or 3 effective digits, in the fitting with numerical results we did not pursue accuracy beyond $1\%$ and the best fit free parameters presented in Table \ref{FSs0} to \ref{CDs2} also have no more than 3 significant digits. 

One uncertainty in the setup of the ejecta density profile is the core radius ratio, i.e. $w_{core}=R_{core}/R_{ej}$. In the numerical simulation, when $n<3$, $w_{core}=0$ is assumed for simplification. When $n>3$, a flat core is assumed to ensure the ejecta have a finite mass.  $w_{core}$ can be estimated by investigating the core velocity to ejecta velocity ratio, i.e. $v_{core}/v_{ej}=w_{core}$. For $n>5$, typical values of $v_{core}$ are found to be between $10^3\, \rm \kms$ and $10^4\, \rm \kms$ \citep{CF94}. $v_{ej}$ of several $10^4\, \rm \kms$ has been observationally measured. Meanwhile $v_{ej}$ should be smaller than the speed of light. So a value between $0.01$ and $0.1$ would be reasonable for $w_{core}$. Here we adopt the values $w_{core}=0.05$ for $s=0$ and $w_{core}=0.1$ for $s=2$, which are slightly larger than the values used in TM99. A larger $w_{core}$ is used for the $s=2$ cases mainly due to numerical considerations. When $s=2$ the density contrast between the ejecta and ambient medium is very high especially for large $n$, which causes difficulty for numerical simulations with very small $w_{core}$. 

A change in $w_{core}$ can affect $\lambda_b$ and $\lambda_c$ in the FE solution when $3<n<5$, plus $\zeta_b$ and $\zeta_c$ in the SSDW solution. When $n\geq 7$, $\zeta_b$ and $\zeta_c$ presented in Tables \ref{FSs0} to \ref{CDs2} for either $w_{core}=0.05$ or $w_{core}=0.1$ are consistent with the asymptotic value at $w_{core}\rightarrow 0$ within $2\%$. For $n=6$, the difference is slightly larger and about $5\%$. It is expected that a choice of different $w_{core}$ within the range $0.1-0.01$ would affect $\zeta_b$ and $\zeta_c$ by only a few percents. Variation of $w_{core}$ does change $\lambda_b$ and $\lambda_c$ with $3<n<5$ significantly, as $\lambda_b \propto \lambda_c \propto w_{core}^{-1}$ when $w_{core}\rightarrow 0$. However,  $3<n<5$ cases are not very important for the study of SNR evolution. Overall we believe different choices of $w_{core}$ within $0.01-0.1$ would not affect the application of our approximate solutions for shock evolution significantly.  


\begin{table}
\centering
\caption{Basic parameters for the analytical approximation of the forward shock in a uniform medium with $s=0$}
\begin{threeparttable}
\begin{tabular}{lccccccl}
\hline\hline
n &$\alpha$&$\zeta_b$&$t^*_{tran}$&$R^*_{tran}$&$|\Delta R_b^*|/R_b^* $\\
\hline

0&1.25&- &0.4 &0.6 &$\lesssim 2\%$\\
1&1.19&- &0.34 &0.56 &$\lesssim 2\%$\\
2&1.10&- &0.24 &0.48 &$\lesssim 2\%$\\
4&0.80&- &0.04&0.2 &$\lesssim 2\%$\\
6&36.1&1.06 &2.29&1.57 &$\lesssim 4\%$\\
7&20.3&1.06 &1.62 &1.35 &$\lesssim 2\%$\\
8&14.7&1.08 &1.33 &1.23 &$\lesssim 2\%$\\
9&10.4&1.12 &1.11 &1.12 &$\lesssim 2\%$\\
10&8.91&1.15 &1.0 &1.07&$\lesssim 2\%$\\
12&7.11&1.21 &0.87 &0.99&$\lesssim 2\%$\\
14&6.23&1.26 &0.79 &0.94&$\lesssim 2\%$\\
\hline\hline
\end{tabular} 
\label{FSs0}
\begin{tablenotes}
\small
\item $w_{core}=0$ for $n<3$ and $w_{core}=0.05$ for $n>3$.
\end{tablenotes}
\end{threeparttable}
\end{table}

\begin{table}
\centering
\caption{Basic parameters for the analytical approximation of the forward shock in a wind profile medium with $s=2$}
\begin{threeparttable}
\begin{tabular}{lccccl}
\hline\hline
n &$\alpha$ &$\zeta_b$&$t^*_{tran}$&$R^*_{tran}$&$|\Delta R_b^*|/R_b^* $\\
\hline
0&0.95&- &0.05 &0.07 &$\lesssim 1\%$\\
1&0.91&- &0.04 & 0.06&$\lesssim 1\%$\\
2&0.85&- &0.02 &0.04 &$\lesssim 1\%$\\
4&0.63&- &0.002 &0.007 &$\lesssim 1\%$\\
6&11.3&0.77 &1.2 &0.83 &$\lesssim 1\%$\\
7&8.00&0.83 &0.64 &0.53&$\lesssim 1\%$\\
8&6.22&0.90 &0.43 &0.4 &$\lesssim 1\%$\\
9&5.16&0.97 &0.32 &0.32 &$\lesssim 1\%$\\
10&4.56&1.03 &0.27 &0.28 &$\lesssim 1\%$\\
12&3.81&1.14 &0.2 &0.22 &$\lesssim 1\%$\\
14&3.40&1.23 &0.16 &0.19 &$\lesssim 1\%$\\

\hline\hline
\end{tabular} 
\label{FSs2}
\begin{tablenotes}
\small
\item $w_{core}=0$ for $n<3$ and $w_{core}=0.1$ for $n>3$.
\end{tablenotes}
\end{threeparttable}
\end{table}

\begin{table}
\centering
\caption{Basic parameters for the analytical approximation of CD in a uniform medium with $s=0$}
\begin{threeparttable}
\begin{tabular}{lcccccl}
\hline\hline
n &a &b &c &$\zeta_c$&$t^*_{lim}$ &$|\Delta R_c^*|/R_c^* $\\
\hline
0&0.89 &-0.37& 1.66&- &4 &$\lesssim 4\%$\\
1&0.94 &-0.25&1.42&- &4.5 &$\lesssim 4\%$\\
2&1.08 &-0.06&1.05 &-&5 &$\lesssim 2\%$\\
4&1.38 &0.27&0.72&- &4 &$\lesssim 5\%$\\
6&5.54 &-0.1&1.11 &0.84 &4.5 &$\lesssim 1\%$\\
7&4.28 &-0.11&1.09 &0.89 &4.5 &$\lesssim 2\%$\\
8&3.09 &-0.16&1.16 &0.94 &4 &$\lesssim 2\%$\\
9&3.01 &-0.13&1.09&0.98 &4 &$\lesssim 3\%$\\
10&2.92 &-0.12&1.07&1.01 &4 &$\lesssim 3\%$\\
12&2.35 &-0.16&1.11&1.08 &4 &$\lesssim 3\%$\\
14&2.20 &-0.16&1.10&1.13 &4 &$\lesssim 4\%$\\
\hline\hline
\end{tabular} 
\label{CDs0}
\begin{tablenotes}
\small
\item $w_{core}=0$ for $n<3$ and $w_{core}=0.05$ for $n>3$.
\end{tablenotes}
\end{threeparttable}
\end{table}

\begin{table}
\centering
\caption{Basic parameters for the analytical approximation of CD in a wind profile medium with $s=2$}
\begin{threeparttable}
\begin{tabular}{lcccccl}
\hline\hline
n &a&b &c &$\zeta_c$&$t^*_{lim}$&$|\Delta R_c^*|/R_c^* $\\
\hline
0&1.23 &0.50&0.53&- &16&$\lesssim 1\%$\\
1&1.16 &0.50&0.54 &- &16&$\lesssim 1\%$\\
2&1.06 &0.51&0.52 &- &16&$\lesssim 1\%$\\
4&0.58 &0.47&0.71 &- &16&$\lesssim 2\%$\\
6&6.04 &0.53&0.47 &0.56 &16&$\lesssim 1\%$\\
7&4.61 &0.53&0.47 &0.64 &16&$\lesssim 1\%$\\
8&3.81 &0.53&0.47 &0.71 &16&$\lesssim 1\%$\\
9&3.35 &0.53&0.47&0.77 &16&$\lesssim 1\%$\\
10&2.94 &0.53& 0.47&0.83 &16&$\lesssim 1\%$\\
12&2.47 &0.52&0.48&0.93 &16&$\lesssim 1\%$\\
14&2.21 &0.52&0.48&1.01 &16&$\lesssim 1\%$\\
\hline\hline
\end{tabular} 
\label{CDs2}
\begin{tablenotes}
\small
\item $w_{core}=0$ for $n<3$ and $w_{core}=0.1$ for $n>3$.
\end{tablenotes}
\end{threeparttable}
\end{table}

\begin{figure}
 \centering
 \includegraphics[width=\columnwidth]{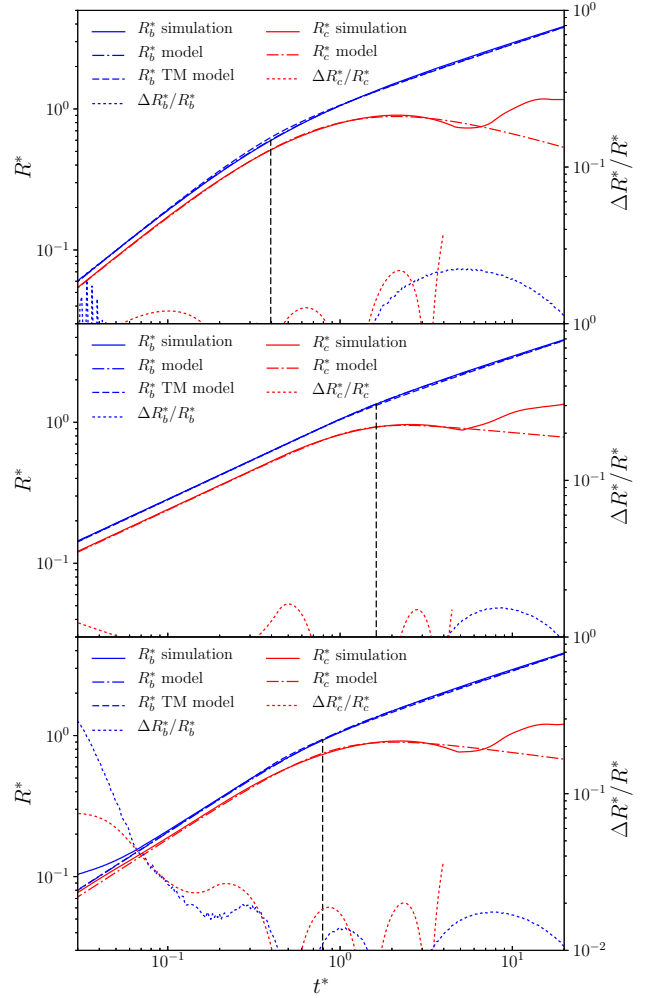} 
\caption{Fitting of the forward shock radius $R_b^*$ and CD radius $R^*_c$ with $s=0$. From top to bottom, the panels  show $n=0$, 7 and 14, respectively. The x-axis is the dimensionless time $t^*$. The left y-axis is the dimensionless radius $R^*$ and the right y-axis is the radius offset $|\Delta R^*|/R^*$. The blue lines are for the forward shock while the red lines are for the CD. The solid line is the simulation result, the dot-dashed line is our model prediction, the dashed line is the TM model estimate and the dotted line is the normalized radius offset between the numerical simulation and our model. The black vertical dashed line characterizes the transition time $t_{tran}^*$  from early FE or SSDW solution to the late ST solution.}  
\label{s0_fit}
 \end{figure}

\begin{figure}
 \begin{center}
 \includegraphics[width=\columnwidth]{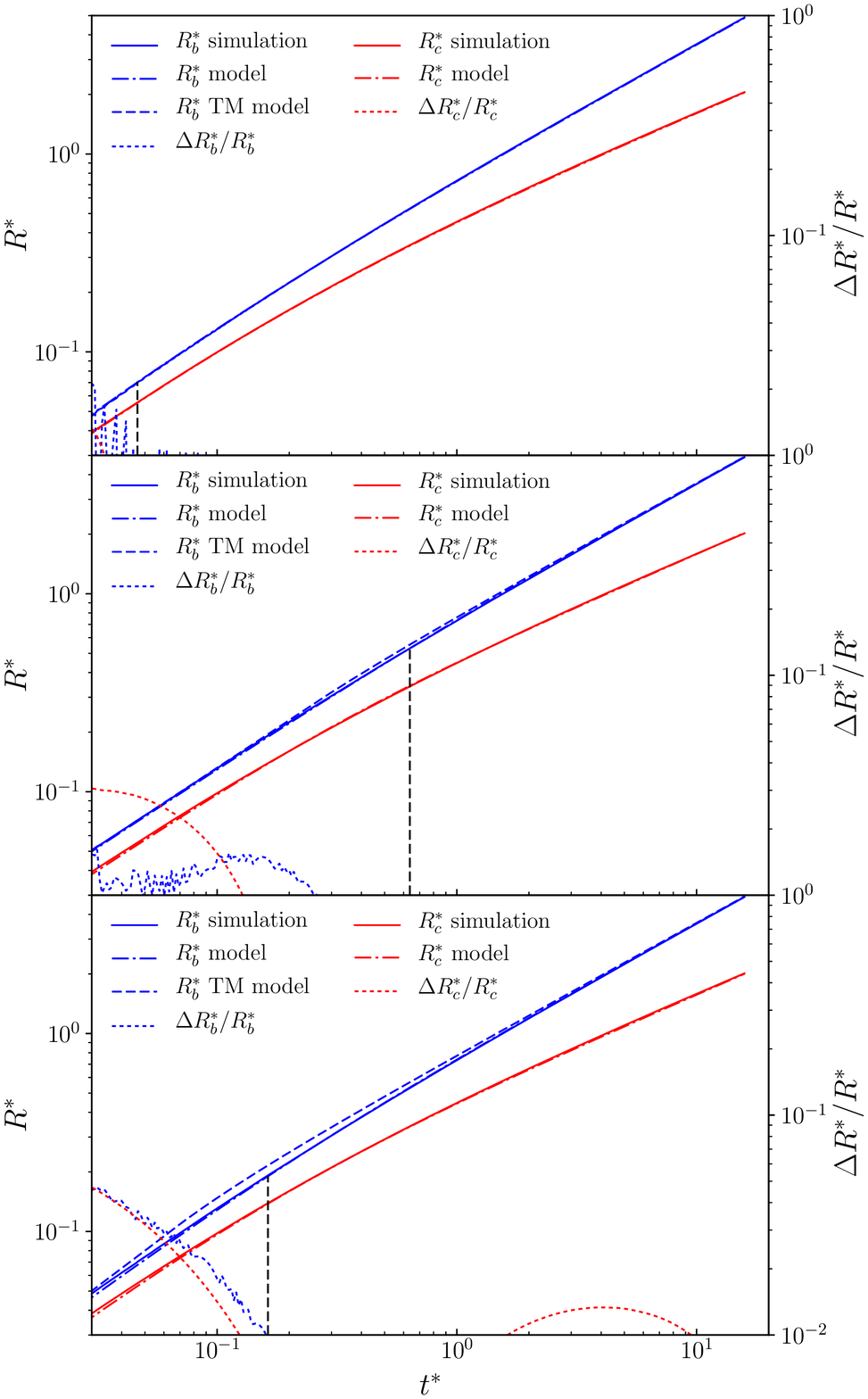} 
  \caption{Same as Fig \ref{s0_fit} but with $s=2$. Note $t_{tran}^*$ in top panel with $n=0$ is very small. } 
    \label{s2_fit}
 \end{center}
 \end{figure}

\subsection{Comparison with numerical simulations and TM model}
In Tables \ref{FSs0} to \ref{CDs2}, we introduce a dimensionless ratio $|\Delta R^*(t^*)| /R^*(t^*)$ to illustrate the performance of our model when comparing with numerical simulations. In $|\Delta R^*(t^*)| /R^*(t^*)$, $R^*$ is the radius estimated from our analytical model and $\Delta R^*$ is the radius offset between the numerical simulation and the analytical approximation. The numerical method applied in this paper is similar as that in TM99 with minor changes; see Appendix \ref{App:numerical_method} for the detailed numerical setup.

In Fig. \ref{s0_fit} and \ref{s2_fit}, we present example fits with selected $n$ for $s=0$ and $s=2$, respectively. The solution from the TM model, if available, is also provided in the figures for comparison. In the figures, we only show the fitting results in the time range $0.03<t^*<20$, which covers the transition region with $t^*\sim t_{tran}^*$ between the early FE solution ($n<5$) or SSDW solution ($n>5$) and the late ST solution. Remnants falling into this time range are found to have a radius between about $0.05R_{ch}$ to $4.5R_{ch}$, where $R_{ch}$ is the characteristic radius defined in section \ref{scale}. For typical SNR parameters, we found the corresponding dimensional radius is roughly in the range $[\rm 0.2pc, 15pc]$ for uniform medium ($s=0$) and $[\rm 0.6pc, 60pc]$ for wind density profile ($s=2$) which should be enough for comparison with observation.  More importantly, the difference between the analytical approximation and simulated remnants is expected to gradually shrink to zero when $t^*$ moves far away from $t^*_{tran}$. $t_{tran}^*$ is shown as a vertical black dashed line in Figs. \ref{s0_fit} and \ref{s2_fit}. In Tables \ref{FSs0} to \ref{CDs2}, we also list the values of $t^*_{tran}$ and $R_{tran}^*$, which characterize the transition time and radius from the early ED phase to the late ST phase.

Next we discuss  our model performance for the forward shock and the CD in detail. We begin with the forward shock model. As shown in Tables \ref{FSs0} and \ref{FSs2}, our model can reproduce the simulation results within about $2\%$ accuracy for $s=0$ and about $1\%$ accuracy for $s=2$. In Fig. \ref{s0_fit} and \ref{s2_fit}, the ratios $|\Delta R_b^*(t^*)| /R_b^*(t^*)$  in the top panel with $n=0$ show spikes at small $t^*$ that are mainly due to the spatial resolution of the simulation. Thus these spikes are not taken into account when counting the uncertainty of our model. For the middle and bottom panels in Figs. \ref{s0_fit} and \ref{s2_fit} with $n>5$, the large offset between the model and numerical simulation at small $t^*$ is mainly introduced by the initial setup of our code. In the simulation for $n>5$, we assume the remnant follows the FE solution at $t_0^*$ when the simulation starts. Although the SSDW solution works like a magnet and the simulated remnant quickly evolves to the SSDW solution. The results at small $t^*$ could still possibly be affected. This is clearly shown in the bottom panel of Fig. \ref{s0_fit}. We found when we decrease $t^*_0$, the deviation at small $t^*$ also decreases. We ignore the deviation at small $t^*$ when deriving the uncertainty $|\Delta R_b^*|  /R_b^*$ shown in Tables  \ref{FSs0} and \ref{FSs2}. 

\begin{figure}
 \begin{center}
 \includegraphics[width=\columnwidth]{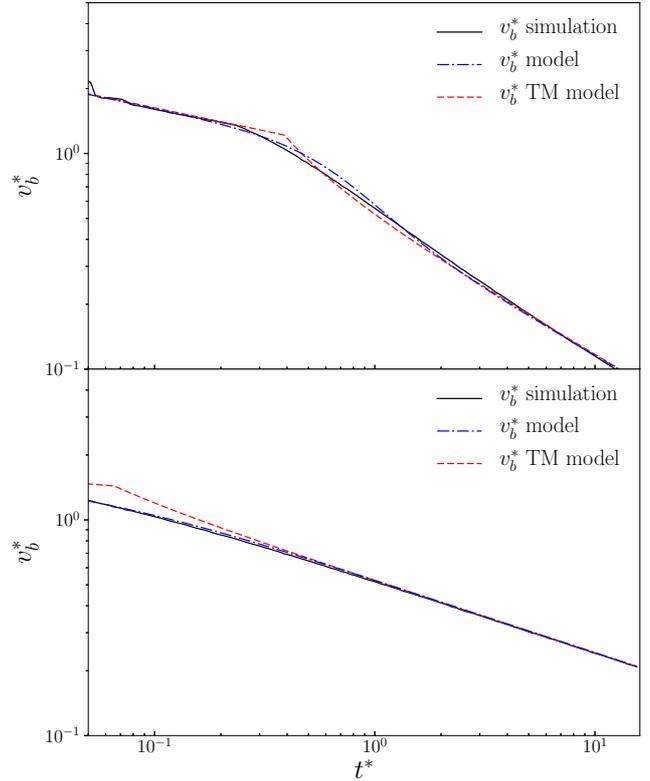} 
  \caption{Dimensionless forward shock velocity as a function of time. The upper panel is for $n=14$ and $s=0$ while the lower panel is for $n=14$ and $s=2$.  The black solid line is the simulation result, blue dot-dashed line is our model prediction, and the red dashed line is the TM model estimate.} 
    \label{forwardvelocity}
 \end{center}
 \end{figure}

According to Figs. \ref{s0_fit} and \ref{s2_fit}, for a uniform medium with $s=0$ both our model and the TM model produce reasonably good fits to the simulations. For a wind density profile with $s=2$, our model provides better performance than the TM model especially for large $n$. It is probably because the solution developed in \cite{Micelotta16} for a wind density profile is simply a connection of the SSDW solution and the {\it general ST solution}. However the SSDW solution is only valid when the reverse shock is still in the ejecta envelope. After the reverse shock enters the flat core, a {\it general ED solution} is needed for the TM  model, especially for $n\gg 5$ cases. Also when \cite{Micelotta16} extend the solutions in TM99 to a wind density profile situation, they do not compare their model results with numerical simulations. The difference between our model and the TM model could be more easily identified in the evolution of velocity. In case the {\it general ED solution} and the {\it general ST} solution are imperfectly connected, the velocity provided by the TM model may exhibit a small break in the transition region while our model velocity is a smooth function of time. For the TM model, we found this break is mainly significant for cases with large $n$. In Fig. \ref{forwardvelocity}, we plot the forward shock velocity as a function of time for two cases, $n=14$, $s=0$ and $n=14$, $s=2$. The simulation velocity presented in the figure is calculated using the strong shock condition, i.e. the shock velocity is  $4/3$ of the post shock velocity. We also calculated the shock velocity with the time derivative, i.e. $v=dR/dt$, as in TM99. We found that the  velocities from the above two methods are consistent with each other for the time range studied here. 

Our model provides a good fit for the $n=4$ case, while the TM model runs into trouble with this case. It is probably due to the following two reasons. At first, when $n=4$ the majority of mass is concentrated in the core of the ejecta while the majority of the energy is stored in the envelope of the ejecta. Because of this special configuration, the remnant can quickly enter the ST phase as long as the bulk of the energy is transferred to the ambient medium, while the majority of the mass in the ejecta still remains unshocked. In our fitting for $n=4$, we found a very small transition time $t_{tran}^*$  (see Tables \ref{FSs0} and \ref{FSs2}) that is much smaller than the $t^*_{ST}$ derived in the TM model. Due to the small $t^*_{tran}$, the remnant simply follows the ST solution during almost the entire time range presented in the figures, which is easy to fit. Secondly our model only depends on the asymptotic behavior of the remnant without any assumption about the dynamical structure. Thus it is not strongly affected by the mass and energy distribution within the ejecta.

We now consider the CD model. As shown in Tables \ref{CDs0} and \ref{CDs2}, our model can reproduce the simulation results within about $4\%$ accuracy for $s=0$ and about $1\%$ accuracy for $s=2$. In Figs. \ref{s0_fit} and \ref{s2_fit}, at small $t^*$  we also see spikes due to limited spatial resolution in the top panel and the large offset due to the initial setup of our code in the middle and bottom panels. As in the forward shock fitting, we ignore those features at small $t^*$ when deriving the uncertainty $|\Delta R_c^*|  /R_c^*$. At large $t^*$, the CD radius exhibits different behaviors in a uniform medium and a wind density profile, so we will discuss  them separately. 

In a uniform medium, the CD radius starts to show oscillations around $t^*\sim 4$, which is probably due to the reflected wave generated when the reverse shock reaches the remnant center. Because of this feature, we only fit the simulation results up to a time $t_{lim}^*$ which marks the beginning of the oscillation phase for the CD. The choices of $t_{lim}^*$ for different $n$ are listed in Table \ref{CDs0}. The parameters $\alpha$, $b$, $c$ and $|\Delta R_c^*|  /R_c^*$ provided in Table \ref{CDs0} are only valid up to  $t_{lim}^*$. Beyond this limit, the evolution of the CD radius is  complicated and hard to model. 

In a wind density profile, we did not see the oscillation  seen in a uniform medium up to $t_{lim}^* \approx 16$. It is probably because there is less material in the surrounding medium and it takes a longer time for the reverse shock to reach the remnant center and generate the reflected wave. Since the asymptotic behavior of $R^*_c$ at $t^* \rightarrow \infty$ is unclear, the parameters $\alpha$, $b$, $c$ and $|\Delta R_c^*|  /R_c^*$ provided in Table \ref{CDs2} are expected to be only valid up to $t_{lim}^*\approx 16$. At $t^*=t^*_{lim}\approx 16$, we find the forward shock radius $R_{b}$ already reaches about $5R_{ch}$. According to eq. \ref{charact_R}, the wind bubble radius $R_b$, for a case with ejecta mass $M_{ej}=\Msun$, mass loss rate $\dot{M}_w=10^{-5}\Msun yr^{-1}$ and wind velocity $v_w=10 \rm km/s$, is about 65 pc.   

\section{REVERSE SHOCK}\label{sec:RS}
\begin{figure}
 \begin{center}
 \includegraphics[width=\columnwidth]{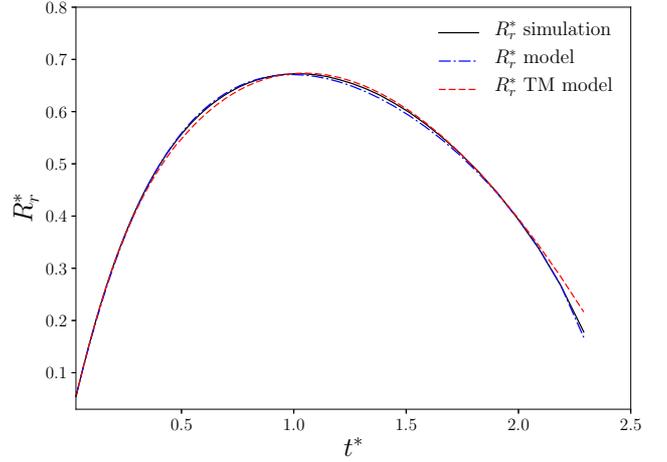} 
  \caption{Reverse shock fitting with $s=0$ and $n=0$ with a linear scale. The black solid line is the simulation result, the blue dot-dashed line is our model prediction and the red dashed line is the TM model estimate.} 
    \label{reverse_shock}
 \end{center}
 \end{figure}

The method developed in this paper depends only on the asymptotic behavior of the remnant and in principal can also be applied to the evolution of the reverse shock radius $R_r^*$. The asymptotic behavior of $R_r^*$  at early times $t\rightarrow 0$ is simply the FE solution for $n<5$ and the SSDW solution for $n>5$. The asymptotic behavior of the reverse shock, when it is approaching the remnant center, however, is not very clear at this point. If we assume the asymptotic behavior of the reverse shock in this limit can be described by the following relation 
\begin{equation}
R_r^*(t^*\rightarrow t_c^*)=c (t_c^*-t^*)^b\rightarrow 0,
\label{Rr_limit}
\end{equation}
where $t_c^*$ represents the dimensionless time when the reverse shock reaches the remnant center. $b$ and $c$ together characterize how fast the reverse shock is approaching the remnant center. The analytical approximation for the reverse shock radius can then be constructed as for the forward shock and CD. One simple way of building the analytical approximation is as follows:
\begin{equation}
\left(\frac{R_r^*}{\lambda_r t^*}\right)^{\alpha}+\left[\frac{R_r^{*}}{c (t_c^*-t^*)^b}\right]^\alpha=1.
\end{equation}
Now we have a solution in the form of $R(t)$ which is 
\begin{equation}
R_r^*=\lbrace(\lambda_r t^*)^{-\alpha}+[c (t_c^*-t^*)^b]^{-\alpha}\rbrace^{-1/\alpha},
\label{Rr_n<5}
\end{equation}
where $\lambda_r(n=0)=\lambda_c(n=0)$ \citep{HS84} is the dimensionless constant for the reverse shock in the FE solution. In Fig. \ref{reverse_shock}, we use the $s=0$ and $n=0$ case as an example to show the validity of the above approximation.  The fitting parameters we use are $\alpha=1.23$, $b=0.58$, $c=0.74$ and $t^*_c=2.37$. The new solution provides a good fit to the numerical simulation and is comparable to results from the TM model. However, because we do not know the asymptotic behavior of the reverse shock when it is approaching the remnant center, we have to make an arbitrary assumption, e.g., eq. (\ref{Rr_limit}). As a result, the approximate solution for the reverse shock now has four free parameters instead of one as for the forward shock, and the model becomes more complicated. Since the overall improvement of eq (\ref{Rr_n<5}) compared to the TM model is not very significant, we do not investigate these solutions. Instead, we recommend that readers  use the TM model solution for the evolution of reverse shock. In Tables \ref{TM_reverse1} and \ref{TM_reverse2}, we summarize the reverse shock solution from the TM model for different situations \citep{TM99,TM2000,HL12,Micelotta16}. 

\begin{table}
\centering
\caption{TM model solution for the reverse shock radius $R_r^*$ with $0\leq n<3$ and $s=0$}
\begin{threeparttable}
\begin{tabular}{l}
\hline\hline
\multicolumn{1}{c}{$t^*<t^*_{ST}$}\\
\hline
\parbox{3cm}{
\begin{equation*}
t^*(R_r^*)=0.707 h R^*_r[1-0.762(3-n)^{1/2}R_r^{*3/2}]^{-2/(3-n)}
\end{equation*}}\\
\hline
\multicolumn{1}{c}{$t^*\geq t^*_{ST}$}\\
\hline
\parbox{3cm}{
\begin{eqnarray*}
R^*_r(t^*)&=& t^*\lbrace 1.56 h^{-1}R^*_{r,ST}-(0.106-0.128n)(t^*-0.639h)\\
&-&\left[\tilde{v}^*_{r,ST} -(0.0676-0.0819n)h\right]\rm ln(1.56h^{-1}t^*)\rbrace
\end{eqnarray*}}\\
\hline
\parbox{3cm}{
\begin{equation*}
h=\left(\frac{3-n}{5-n}\right)^{1/2}
\end{equation*}}\\ 
\hline\hline
\end{tabular} 
\label{TM_reverse1}
\begin{tablenotes}
\small
\item $t^*_{ST}$, $R^*_{r,ST}$ and $\tilde{v}^*_{r,ST}$ are given in Table 3 of TM99.
\item Reference: \cite{TM99} and \cite{TM2000}.
\end{tablenotes}
\end{threeparttable}
\end{table}

\begin{table}
\centering
\caption{TM model solution for the reverse shock radius $R_r^*$ with $5< n\leq 14$ and $s=0,2$}
\begin{threeparttable}
\begin{tabular}{l}
\hline\hline
\multicolumn{1}{c}{$t^*<t^*_{core}$}\\
\hline
\parbox{3cm}{
\begin{equation*}
R_r^*=\frac{1}{l_{ED}}\left\lbrace v_{core}^{*n-3}\frac{(3-s)^2}{n(n-3)}\frac{3}{4\pi}\frac{l^{n-2}_{ED}}{\phi_{ED}}\right\rbrace^{1/(n-s)} t^{*\frac{n-3}{n-s}}, 
\end{equation*}}\\
\hline
\multicolumn{1}{c}{$t^* \geq t^*_{core}$}\\
\hline
\parbox{3cm}{
\begin{equation*}
R^*_r=\left[ \frac{R^*_b(t^*=t_{core}^*)}{l_{ED}t^*_{core}}-\frac{3-s}{n-3}\frac{v^*_b(t^*=t^*_{core})}{l_{ED}}\rm ln\frac{t^*}{t^*_{core}}\right]t^*, 
\end{equation*}}\\
\hline
\parbox{3cm}{
\begin{eqnarray*}
t^*_{core}&=&\left[\frac{l^{s-2}_{ED}}{\phi_{ED}}\frac{3}{4\pi}\frac{(3-s)^2}{n(n-3)}\right]^{1/(3-s)}\frac{1}{v_{core}^*}\\
v^*_{core}&=&\left[\frac{10(n-5)}{3(n-3)}\right]^{1/2}\\
l_{ED}&=&1+\frac{8}{n^2}+\frac{0.4}{4-s}\\
\phi_{ED}&=& [0.65-\exp (-n/4)]\sqrt{1-\frac{s}{3}}
\end{eqnarray*}}\\
\hline\hline
\end{tabular} 
\label{TM_reverse2}
\begin{tablenotes}
\small
\item Reference: \cite{Micelotta16}.
\end{tablenotes}
\end{threeparttable}
\end{table}

\section{Discussion and Summary}{\label{sec:DS}}
In  fitting the CD, we assume that the CD asymptotically approaches the power law relation $R_c^* \sim ct^{*b}$. For a wind density profile ($s=2$), the values of $b$ and $c$  (see Table \ref{CDs2}) we found are almost constant for ejecta with different density profiles, which implies a universal asymptotic limit for the CD like the ST solution for the forward shock. For a uniform medium, the derived $b$ and $c$ show larger variations because the reflected wave driven by the reverse shock complicates the situation. Thus we have to apply an arbitrary upper cutoff $t_{lim}^*$ during the fitting, which could affect the values of $b$ and $c$. 


In this paper, we present a new  approach to derive analytical approximations describing the shock evolution in a non-radiative SNR. The new approach depends on only the asymptotic behaviors of the remnant during its evolution and thus is greatly simplified compared with the TM model. We then use the new method to closely investigate the shock evolution in a non-radiative SNR in both the interstellar medium with a constant density profile and a circumstellar medium with  a wind density profile. We focus on the study of the forward shock and CD while application to the reverse shock is also briefly discussed.  We compare our new analytical approximation with numerical simulations  and find that a few percent accuracy is achieved for all investigated cases. For the forward shock, we also compare our new solutions to the TM model. In a uniform ambient medium, our solutions are comparable to the TM model while for a wind density profile medium our solutions perform better, especially when the ejecta envelope has a steep density profile. In order to obtain the analytical solution, we made several simplifying  assumptions. Possible extensions of the current solutions to more complicated situations will be studied in future work. The transition from the ST phase to the radiative phase in SNR evolution has been discussed in \cite{Cioffi88}. In the future, we would like to use the method developed here to investigate the problem. 

\section*{Acknowledgements}
We would like to thank the referee Dr. Christopher F. McKee for useful comments and constructive suggestions that helped us  improve the manuscript. We would also like to thank Dr. Fabio Acero for discussion about the forward shock velocity. XT got the main idea for this work while undertaking PhD research with RAC at the Department of Astronomy University of Virginia and then moved to Max Planck Institute for Astrophysics to finish most of the work. XT would like to thank the Department of Astronomy at UVa for a stimulating atmosphere, and Eugene Churazov and Rashid Sunyaev at MPA for support of this work.  The research was supported in part by NASA grant NNX112AF90G.





\appendix
\section{Dimensionless constants for asymptotic solutions}{\label{App:basic_params}}
The density profile applied here is the same as that in TM99:
\begin{equation}
  \rho(r,t)=\begin{cases}
    \rho_{ej}(r)=\frac{M_{ej}}{R_{ej}^3}f(\frac{r}{R_{ej}}), & r\le R_{ej}\\
    \rho_a(r) = \eta_s r^{-s}  & r> R_{ej},
  \end{cases}
  \label{density}
\end{equation}
where $R_{ej}$ is the radius of the outer boundary of the ejecta and $\eta_s$ is a constant. $f(r/R_{ej})$ is the structure function of the ejecta. For freely expanding ejecta, we assume the following core-envelope power law profile:
\begin{equation}
  f(w)=\begin{cases}
    f_0, & 0\le w \le w_{core}\\
    f_0 (w_{core}/w)^{n}  & w_{core}\le w \le 1  ,
  \end{cases}
  \label{density_ej}
\end{equation}
where $w=r/R_{ej}$ and $w_{core}=R_{core}/R_{ej}$.  Since the total mass of ejecta is assumed to be $M_{ej}$, we obtain
\begin{equation}
f_0=\frac{3}{4\pi w_{core}^n}\left[\frac{1-(n/3)}{1-(n/3)w^{3-n}_{core}}\right].
\label{f0}
\end{equation}
Throughout the paper, $s<3$ is required to ensure a finite mass of the swept up ambient medium, 
 
With the above density distribution, $\lambda_b$ and $\lambda_c$ in the FE solution can be derived and expressed explicitly. Based on energy conservation, we have
\begin{equation}
E_{SN}= \frac{1}{2}\int_0^{R_{ej}}4\pi r^2 \rho_{ej}(r) \left(\frac{r}{t}\right)^2dr.
\end{equation}
Since in the FE solution $R_{ej}=R_c=\lambda_c t\sqrt{E_{SN}/M_{ej}}$, after some calculation we obtain 
\begin{equation}
\lambda_c^2(n,w_{core})=2w_{core}^{-2}\left(\frac{5-n}{3-n}\right)\left( \frac{w_{core}^{n-3}-n/3}{w_{core}^{n-5}-n/5} \right),
\label{v_free}
\end{equation}
which is consistent with eq. (27) in TM99. When $n<3$, a core is not necessary, so we assume $w_{core}=0$ for $n<3$. $\lambda_c$ now simply becomes
\begin{equation}
\lambda_c^2(n<3)=2\left(\frac{5-n}{3-n}\right)
\end{equation}
According to the discussion in \cite{Parker63} and \cite{HS84}, in the FE solution
\begin{equation}
\lambda_b=q_b\lambda_c
\end{equation} 
where $q_b=1.1$ for a uniform ambient medium with $s=0$ and $q_b=1.19$ for a wind profile medium with $s=2$.

For an ideal gas with specific heat index $\gamma=5/3$, the dimensionless constant $\xi(s)$ defined in the ST solution equals  $2.026$ when $s=0$ and $3/2\pi$ when $s=2$ \citep{Taylor46,Sedov59,Book94}.
For arbitrary $s$, \cite{O&M} found that the following expression
\begin{equation}
\xi(s)=\frac{(5-s)(10-3s)}{8\pi}
\label{xi}
\end{equation}
provides a good approximation.

By definition, $R_c^*=\zeta_c t^{*(n-3)/(n-s)}$ and $R_b^*=\zeta_b t^{*(n-3)/(n-s)}$ in the SSDW solution. Through comparison with the SSDW solution in \cite{Chevalier82} we found that 
\begin{equation}
\zeta_c=\left(Af_0w_{core}^n\lambda_c^{n-3}\right)^{1/(n-s)}.
\label{zeta_c}
\end{equation}
and
\begin{equation}
\zeta_b=\left(\frac{R_1}{R_c}\right)\zeta_c=\left(\frac{R_1}{R_c}\right)\left(Af_0w_{core}^n\lambda_c^{n-3}\right)^{1/(n-s)}.
\label{zeta_b}
\end{equation}
where $A$ and $R_1/R_c$ are the coefficients provided in Table 1 of \cite{Chevalier82}.

Clearly $\zeta_c$ depends on the value of $w_{core}$. In the limit $w_{core}\rightarrow 0$,
\begin{equation}
\zeta_c=\left\lbrace \frac{3A(n-3)}{4\pi n}\left[ \frac{10}{3}\left( \frac{5-n}{3-n}\right)\right]^{(n-3)/2} \right\rbrace^{1/(n-s)}.
\end{equation}
where we have used eqs.  (\ref{f0}) and (\ref{v_free}). 

\section{Numerical setup}{\label{App:numerical_method}}

We use the one dimensional hydrodynamic code described in  Appendix B of \cite{TM99}. It uses a Lagrangian finite differencing scheme with a standard formulation for artificial viscosity as discussed in \cite{R&M67}. We replaced the artificial viscosity with 
\begin{equation}
q=\rho\left\lbrace c_2 \frac{\gamma+1}{4}|\Delta v|+\sqrt{c_2^2\left(\frac{\gamma+1}{4}\right)^2\Delta v^2+c_1^2c_s^2}\right\rbrace |\Delta v|,
\label{ArtificialViscosity}
\end{equation}
where $\Delta v$ is the velocity jump across a zone, $\rho $ is the density of the zone and $c_s$ is the sound speed in the zone \citep{Caramana98}. Physically, the artificial viscosity described in eq. \ref{ArtificialViscosity} is designed to mimic the Rankine Hugoniot jump conditions in real shocks. If we assume $c_1=c_2=1$ and consider $\Delta v$ as the velocity jump across the shock front, eq. \ref{ArtificialViscosity} then represents the pressure jump at the shock front according to the Rankine Hugoniot jump conditions. In practice, the linear viscosity term in eq. \ref{ArtificialViscosity} with coefficient $c_1$ can damp the spurious oscillations appearing behind the shock front due to the application of a quadratic type viscosity presented in \cite{R&M67}. In our simulation, we assume $c_2=1$ and $c_1=0.5$, which we found is able to damp the spurious oscillations and confine the shock to  a few zones.  A Lagrangian Courant-Friedrichs-Lewy (CFL) condition is utilized in all the simulations with
CFL number of 0.5 and the increase in time step is required to be no more than $5\%$ between steps.

The shock locations are determined by the position of maximum pseudo pressure. In order to achieve effective subzone resolution of the shock positions, we interpolate the pseudo pressure in the region around the shock contact discontinuity with a cubic-spline interpolation as in TM99.

We take the FE solution as the initial setup for all the simulations. For $n>5$ cases, the SSDW solution works like a magnet and the remnant rapidly approaches the SSDW solution in the simulation. However a small starting time $t_0$ is still necessary to ensure accurate results in the parameter range of interest. Thus $t_0^*=3\times 10^{-4}$ is chosen for all the initial setup. 

When $n<3$, a total of 1024 equal mass zones are put in the ejecta. When $n>3$, a flat density core is assumed and is divided into 512 equal mass zones. The number of zones selected for the envelope depends on $n$ and ranges from $10^3$'s to 10$^4$'s to ensure energy conservation within $1\%$.
For $s=0$,  equal size zones are put in the ambient medium between the initial radius $r_0^*=\lambda_c t_0^*$ and the outer boundary $R^*_o=5$. The number of zones again is selected to ensure energy conservation within $1\%$ and ranges from $10^3$'s to 10$^4$'s. For $s=2$, the same number of equal size zones are put in both the region $r_0^*-3r_0^*$ and region $3r_0^*-R^*_o$. The exact number of zones is again selected to ensure energy conservation within $1\%$ for $n>5$ and $0.5\%$ for $n<5$. In the simulation for the reverse shock shown in Fig. \ref{reverse_shock}, we put 1024 equal radial size zones in both the ejecta and ambient medium to obtain accurate positions for the reverse shock.

In the evolution of a non-radiative SNR, we have $R\propto E_{SN}^{m}$ where $1/5\leq m\leq 1/2$. $m$ equals to $1/2$ in the FE solution while $m=1/5$ corresponds to ST solution.  As a result, a $1\%$ offset in energy $E_{SN}$ would in principal result in $m\%\lesssim 0.5\%$ offset in radius $R$. Partly because of this, in  fitting the numerical simulations we did not pursue $|\Delta R|/R $ beyond $1\%$. In practice, we found that when we further increase the spatial resolution to achieve energy conservation better than $1\%$, the variations in the forward shock and CD positions already become negligible.

\bsp	
\label{lastpage}

\end{document}